\documentstyle[psfig,conf_iap]{article}

\def\cwcplotmacro#1#2#3#4#5#6#7{\centering \leavevmode
    \vbox to#2{\rule{0pt}{#2}}
    \includegraphics{#1}}

\begin{document}

\heading{Mg II Selected Absorbers: Ionization Structures and a
Survey of Weak Systems}

\par\medskip\noindent

\author{Christopher W. Churchill$^{1}$}

\address{Department of Astronomy, Pennsylvania State University, 
University Park, 16802}

\begin{abstract}
First results from a study of high ionization absorption properties in
$\sim 30$ Mg~{\sc ii} absorption selected galaxies are presented.
We have tested for correlations of Mg~{\sc ii}, C~{\sc iv},
Si~{\sc iv}, N~{\sc v}, and O~{\sc vi} equivalent widths 
with the galaxy properties and Mg~{\sc ii} gas kinematics.
The results are suggestive of multi--phase halos with little to
no global ionization gradient with impact parameter.
C~{\sc iv} may arise in both the Mg~{\sc ii}--Ly$\alpha$ clouds
and a high ionization ``halo'' traced by O~{\sc vi}.
We also report on an unbiased survey for weak Mg~{\sc ii} systems
using HIRES/Keck spectra.
At $\left< z \right> = 0.9$, we find $dN/dz = 1.6\pm0.1$
for $0.02 < W(\lambda 2796) < 0.3$~{\AA} and that
weak systems comprise $\sim 65$\% of all Mg~{\sc ii} absorbers.
In all but one case, the weak Mg~{\sc ii} systems exhibit Fe~{\sc ii}
absorption with $\left< \log N(\hbox{Fe~{\sc ii}})/N(\hbox{Mg~{\sc
ii}}) \right> = -0.3\pm 0.4$ measured for the sample.  
We suggest that weak Mg~{\sc ii} absorbers comprise a substantial
yet--to--be explored population. 
If weak systems select the LSB and/or dwarf galaxy population, 
then the weakest Mg~{\sc ii} absorbers may provide one of the
most sensitive tracers of chemical enrichment and evolution of the  UV
background from $z=2$ to $z=0$.
\end{abstract}

\section{Mg II Selected Galaxies and Absorbers}

Our understanding of Mg~{\sc ii} absorbers and their relationship to
normal field galaxies has grown considerably in recent years 
\cite{cwcref:bergeron,cwcref:steideleso,cwcref:csv}.
Yet there remain controversial and unanswered questions regarding
detailed {\it cause and effect\/} relationships between gas seen
in absorption and galaxies seen in emission.
How do absorbing gas chemical, ionization, and kinematic conditions
depend upon and influence star formation histories, merging events,
and luminous morphologies of associated galaxies?
It is hoped clues to this question can be extracted from observed
trends, or non--trends, of the low and high ionization gas properties
with the galaxy properties and/or gas kinematics. 

The largest uniform survey of Mg~{\sc ii} systems is complete
to $W(\lambda 2796) = 0.3$~{\AA} \cite{cwcref:ss92}.
The number density of the very weakest Mg~{\sc ii} absorbers and their
chemical and ionization conditions remain unexplored.
What is their number density, what class of {\it luminous\/} object
does $W(\lambda 2796) \sim 0.1$~{\AA} Mg~{\sc ii} absorption 
select, and what astrophysics can we extract from these systems?

In this contribution we report first results from (1) an HST Archival
study of the neutral hydrogen and high ionization absorption lines of
galaxies for which the luminous properties and Mg~{\sc ii} gas
kinematics are measured, and (2) an unbiased survey of $W(\lambda
2796) < 0.3$~{\AA} systems.

\section{Ionization Structure of $W(\lambda 2796) > 0.3$~{\AA} Systems}

We have compiled a database of $\sim 30$ Mg~{\sc ii} absorption
selected galaxies by combining the galaxies of Steidel, Dickinson, \&
Persson \cite{cwcref:sdp}, the HIRES Mg~{\sc ii} absorption profiles
of Churchill, Vogt, \& Charlton \cite{cwcref:cvc} and Archival
FOS/HST spectra of neutral hydrogen and high ionization absorption
lines.
The FOS spectra reduction and line identification methods were
identical to those of the QSO Absorption Line Key Project
\cite{cwcref:keyproj,cwcref:don}.

We tested for correlations of ionization conditions and strengths of
the absorbing gas with: 
(1)~Mg~{\sc ii} velocity spread, $\omega_{v}$,
(2)~galaxy $B$ and $K$ magnitudes and $B-K$ colors, and
(3)~projected galactocentric distance, $D$.
In Table~1, a selection of Kendall $\tau$ rank correlation tests are
tabulated.  
Columns 3, 4, and 5 are the number of data points, the Kendall $\tau$
and the significance level that the tested quantities are not
uncorrelated in units of $\sigma$.

\begin{center}
\begin{tabular}{llrrrl}
\multicolumn{5}{l}{{\bf Table 1.} Rank Correlation Tests} \\
\hline \hline
Property 1 & Property 2 & $N$ & $\tau_{\rm k}$ & $N(\sigma)$ & Notes \\
\hline
$W(\hbox{C~{\sc iv}})/W(\hbox{Mg~{\sc ii}})$ & $W(\hbox{Mg~{\sc ii}})$  & 19 & $-0.65$ & 3.86 \\
$W(\hbox{Mg~{\sc ii}})$                      & $W(\hbox{Ly}~\alpha)$     & 19 & $ 0.48$ & 2.85 \\
$W(\hbox{S~{\sc iv}})/W(\hbox{C~{\sc iv}})$  & $W(\hbox{C~{\sc iv}})$   & 12 & $-0.58$ & 2.65 \\
$W(\hbox{C~{\sc iv}})/W(\hbox{Mg~{\sc ii}})$ & $W(\hbox{Ly}~\alpha)$     & 14 & $-0.53$ & 2.64 \\
$W(\hbox{O~{\sc vi}})$                       & $W(\hbox{C~{\sc iv}})$   &  6 & $ 0.87$ & 2.44 \\
$W(\hbox{C~{\sc iv}})$                       & $W(\hbox{Mg~{\sc ii}})$  & 19 & $ 0.21$ & 1.23 \\
$W(\hbox{C~{\sc iv}})$                       & $W(\hbox{Ly}~\alpha)$     & 19 & $ 0.06$ & 0.28 \\
\hline
$W(\hbox{C~{\sc iv}})$                       & $\omega_{v}(\hbox{Mg~{\sc ii}})$ & 17 & $ 0.61$ & 3.43 \\
$W(\hbox{Mg~{\sc ii}})$                      & $\omega_{v}(\hbox{Mg~{\sc ii}})$ & 30 & $ 0.27$ & 2.12 \\
$W(\hbox{O~{\sc vi}})$                       & $\omega_{v}(\hbox{Mg~{\sc ii}})$ &  8 & $ 0.29$ & 0.99 \\
\hline
$W(\hbox{Mg~{\sc ii}})$                      & $D$ & 19 & $-0.30$ & 1.78 & 1 pt dominates\\
$W(\hbox{C~{\sc iv}})/W(\hbox{Mg~{\sc ii}})$ & $D$ & 11 & $ 0.40$ & 1.73 & 1 pt dominates\\
$W(\hbox{Ly}~\alpha)$                        & $D$ &  8 & $-0.36$ & 1.24 & 1 pt dominates\\
$W(\hbox{C~{\sc iv}})$                       & $D$ & 11 & $ 0.11$ & 0.47 \\
\hline
\end{tabular}
\end{center}

\begin{figure}[t]
\cwcplotmacro{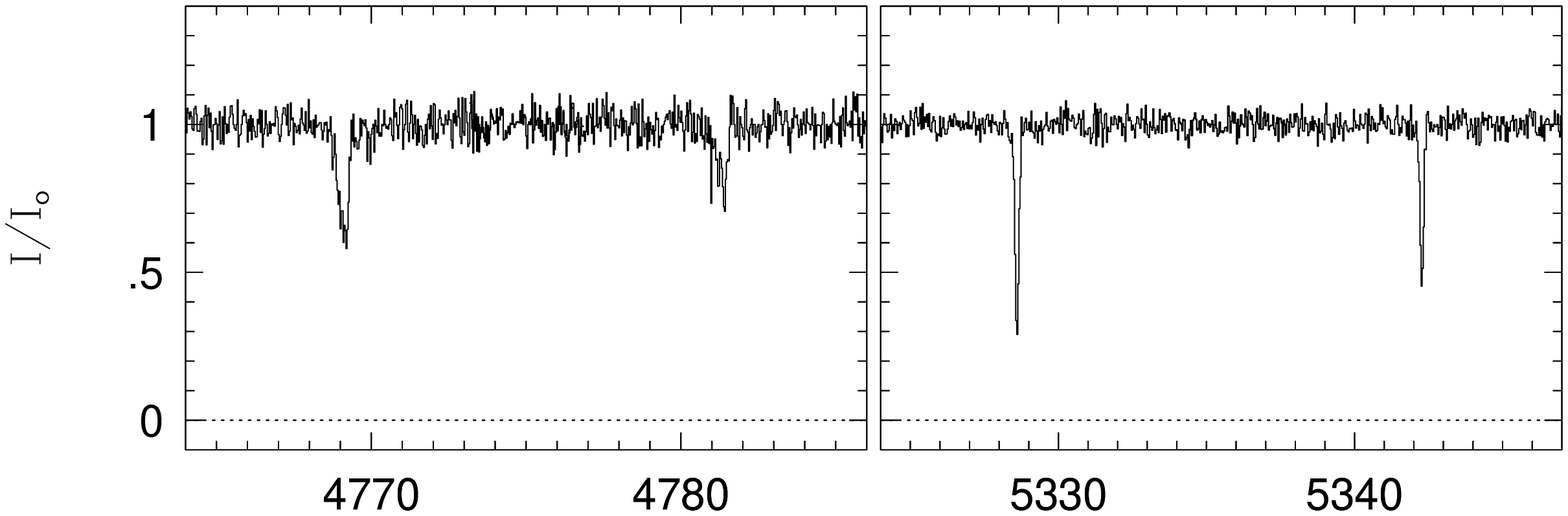}{2.5in}{0}{48.}{40.}{-179}{-8}
\cwcplotmacro{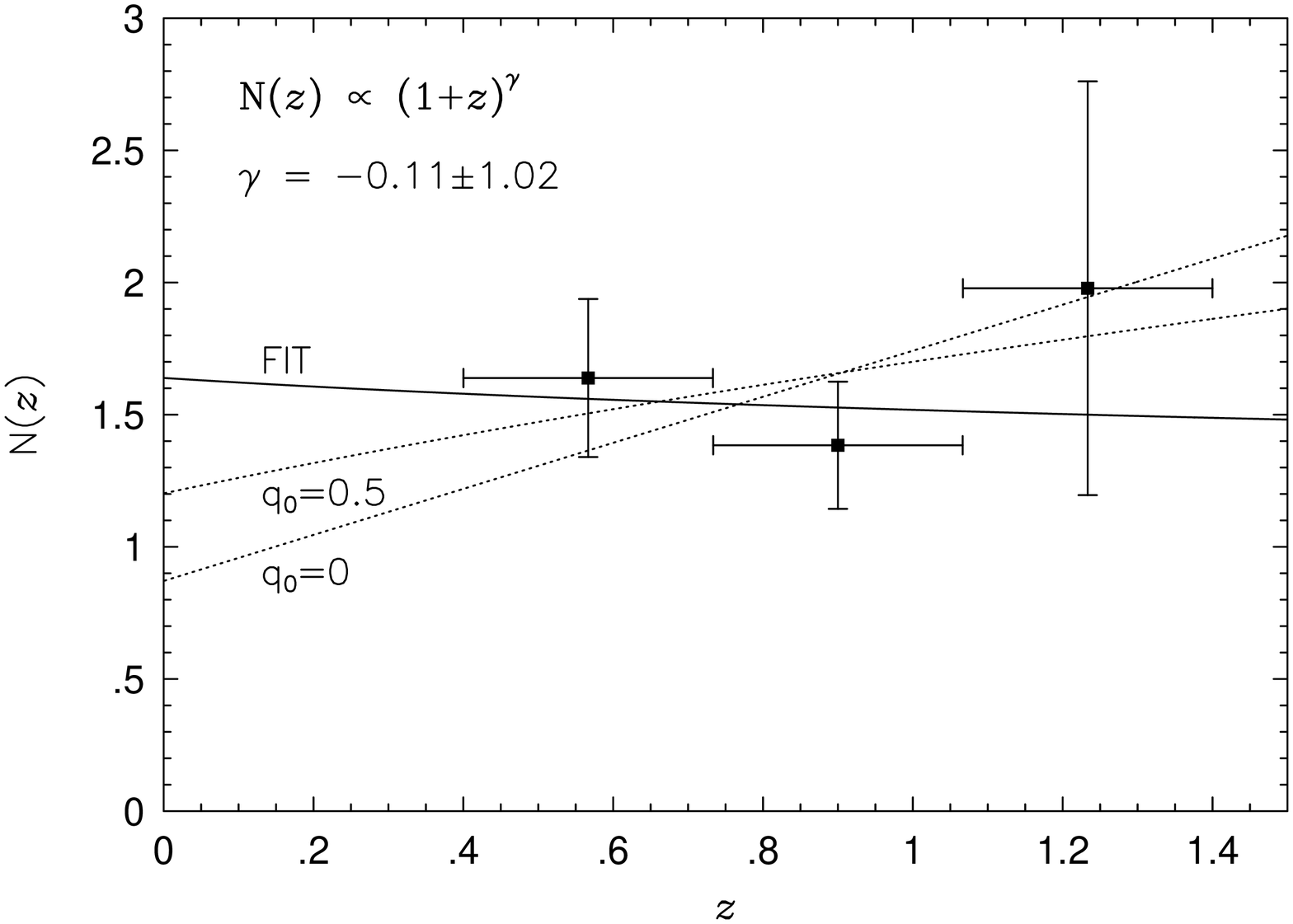}{3.5in}{0}{48.}{40.}{-179}{155}
\vglue -2.58in
\caption[]{(upper) Example of weak Mg~{\sc ii} absorbers. --- (lower)
The number density of $W(2796) < 0.3$~{\AA} absorbers from a redshift
pathlength of $\sim 15$.  The dotted lines show the slopes of
the best fit no--evolution models.}
\end{figure}

The correlation tests are suggestive of multi--phase ionization
with C~{\sc iv} present in both phases.
The clouds of Mg~{\sc ii} trace the neutral hydrogen (Ly$\alpha$) in
the galaxies and have velocity spreads of $\sim 70$~km~s$^{-1}$.
The clouds may be embedded in an extended ``halo'' traced by
O~{\sc vi} absorption \cite{cwcref:bergeron}.
$W(\hbox{C~{\sc iv}})$ does not trace $W(\hbox{Mg~{\sc ii}})$ but
appears to trace $W(\hbox{O~{\sc vi}})$.
We infer that a significant fraction of the C~{\sc iv} is spatially
distributed with the putative O~{\sc vi} halo.
However, the $W(\hbox{C~{\sc iv}})$ is strongly correlated with the
Mg~{\sc ii} velocity spread, whereas $W(\hbox{O~{\sc vi}})$ is not.
This suggests many Mg~{\sc ii} clouds have ionization structures 
in which C~{\sc iv} surrounds their lower ionization
Ly$\alpha$--Mg~{\sc ii} cores. 

There is no evidence for correlations above the 1.5$\sigma$ level
of absorption strengths, $W(\hbox{X})$, and ionization levels,
$W(\hbox{X}_{1})/W(\hbox{X}_{2})$, with the galaxy magnitudes and
colors.
Apparently, for $W(\lambda 2796) > 0.3$~{\AA} selected systems, the 
stellar population is not strongly interfaced with the absorption
properties.
There is no compelling evidence for strong correlations of absorption
properties with projected galactocentric distance, $D$.
Perhaps the spatial distribution of Mg~{\sc ii} clouds is independent
of galactocentric distance and there is no {\it smooth\/} ionization
gradient surrounding these galaxies.
A study that includes the galaxy morphologies, line of sight
orientations, and stellar velocities may reveal a yet--unknown
interrelationship between absorbing gas and galaxies.

\section{The Properties of $W(\lambda 2796) < 0.3$~{\AA} Systems}

Because the lines of sight studied with the HIRES spectra were
selected based upon the presence of known $W(\lambda 2796) > 0.3$~{\AA}
absorbers \cite{cwcref:ss92}, the number density of these systems is
biased too high in our sample.
However, the total redshift path is unbiased for $W(\lambda 2796) <
0.3$~{\AA} systems.
A systematic search for weak Mg~{\sc ii} doublets has been performed
for a redshift path of $\sim 15$ over $0.4<z<1.4$ complete to $W(\lambda
2796) =  0.02$~{\AA} (5$\sigma$) \cite{cwcref:weak}. 
Following Lanzetta et~al.~\cite{cwcref:lanzetta}, the number density
per unit redshift was computed.

In Figure~1, the number density of $W(\lambda 2796)
< 0.3$~{\AA} systems is shown.
We find $dN/dz = 1.6\pm0.1$ at $\left< z \right> = 0.9$.  
Accounting for $dN/dz = 0.9\pm0.1$ found for $W(\lambda 2796) > 0.3$~{\AA}
systems by Steidel \& Sargent \cite{cwcref:ss92}, the number density
of Mg~{\sc ii} systems complete to $W(\lambda 2796) = 0.02$~{\AA} is
$2.5\pm 0.2$ at $\left< z \right> = 0.9$.
The $W(\lambda 2796) < 0.3$~{\AA} systems comprise $\sim
65$\% of all Mg~{\sc ii} absorbing systems at low redshift.
For the sample, $\left< \log N(\hbox{Fe~{\sc ii}})/N(\hbox{Mg~{\sc
ii}}) \right> = -0.3\pm 0.4$.

Applying the maximum likelihood technique to the parameterized relation 
$dN/dz \equiv N(z) = N_0 (1+z)^{\gamma}$, we obtained $\gamma =
-0.1\pm1.0$.
This poorly constrained value is not inconsistent with a non--evolving
population.
A larger sample with extended redshift coverage to $z>2$ would allow
the redshift evolution of these weakest systems to be tested.
We suggest two types of evolution are expected over $0.4<z<2.2$.
First, $dN/dz$ likely increases with decreasing redshift as the
ionizing UV background becomes less intense.  
Second, the ratio of $N(\hbox{Fe~{\sc ii}})/N(\hbox{Mg~{\sc ii}})$
should increase with decreasing redshift for two reasons, provided
the cloud densities do not strongly evolve.  (1) For UV background
photoionization, the higher $J_0$ at higher $z$ produces a smaller
ratio.  
This is consistent with the observed $W(\hbox{C~{\sc
iv}})/W(\hbox{Mg~{\sc ii}})$ evolution \cite{cwcref:bergeron}.
(2) The chemical build up of iron--group elements (a SN~Ia 
process taking several Gyr) would gradually increase the Fe/Mg
abundance ratios following the $z\sim2$ peak star formation epoch of
the universe in which Type II SNe disperse $\alpha$--group elements
such as silicon and magnesium.

If weak systems are closely associated with LSB and/or dwarf galaxies,
there in fact may be a narrow redshift range at $z\sim1$ over which a
``break'' from UV background to local late--type stellar dominated
photoionization occurs as the UV background falls below a critical
level.
Since iron--group enrichment is gradual, such a break could be
inferred from a rapid jump in the fraction of weak Mg~{\sc ii} systems
having strong Fe~{\sc ii} absorption.
The population of weak Mg~{\sc ii} systems may provide one of the
most sensitive probes of the UV background evolution from $z=2$ to
$z=0$.

\acknowledgements{The National Science Foundation has supported this
work through AST--9617185.   I extend special thanks to my
collaborators J.~Charlton, B.~Jannuzi, S.~Kirhokas, J.~Rigby,
D.~Schneider, and C.~Steidel for encouraging me to present results
prior to their publication.  Thanks to J. Bergeron and J. Charlton for
insightful conversations, and to S.~Vogt for HIRES.}

\vfill
\end{document}